\documentclass[preprint,showpacs,preprintnumbers,amsmath,amssymb]{revtex4}

\usepackage{graphicx}
\usepackage{dcolumn}
\usepackage{bm}

\def\sqr#1#2{{\vcenter{\hrule height.#2pt\hbox{\vrule width.#2pt height#1pt \kern#1pt \vrule width.#2pt}\hrule height.#2pt}}}

\begin{document}


\title{Quantum spacetime fluctuations: Lamb Shift and hyperfine structure of the hydrogen atom}

\author{J. I. Rivas}
 \email{jirs@xanum.uam.mx}\affiliation{Departamento de F\'{\i}sica,
 Universidad Aut\'onoma Metropolitana--Iztapalapa\\
 Apartado Postal 55--534, C.P. 09340, M\'exico, D.F., M\'exico.}

 \author{A. Camacho}
 \email{acq@xanum.uam.mx} \affiliation{Departamento de F\'{\i}sica,
 Universidad Aut\'onoma Metropolitana--Iztapalapa\\
 Apartado Postal 55--534, C.P. 09340, M\'exico, D.F., M\'exico.}

 \author{E. G\"okl\"u}
 \email{cbi920000370@xanum.uam.mx} \affiliation{Departamento de F\'{\i}sica,
 Universidad Aut\'onoma Metropolitana--Iztapalapa\\
 Apartado Postal 55--534, C.P. 09340, M\'exico, D.F., M\'exico.}


\date{\today}

\begin{abstract}
We consider the consequences of the presence of metric
fluctuations upon the properties of a hydrogen atom. Particularly,
we introduce these metric fluctuations in the corresponding
effective Schr\"odinger equation and deduce the modifications that
they entail upon the hyperfine structure related to a hydrogen
atom. We will find the change that these effects imply for the
ground state energy of the system and obtain a bound for its size
comparing our theoretical predictions against the experimental
uncertainty reported in the literature. In addition, we analyze
the corresponding Lamb shift effect emerging from these
fluctuations of spacetime. Once again, we will set a bound to
these oscillations resorting to the current experimental outcomes.

\end{abstract}

\maketitle

\section{Introduction}

The quest for a quantum theory of gravity has found several
difficulties, which can be categorized, roughly, as conceptual and
mathematical. Indeed, for instance, string theory has a very large
number ($\sim 10^{523}$) of compactification processes
\cite{becker} . This last remark entails, at least, three
different questions: (i) which procedure, among the existing
possibilities, is the one chosen by nature?; (ii) how and why
nature makes this aforementioned choice?; (iii) how could this
model be tested?

Another, and very popular candidate, loop quantum gravity faces
profound mathematical problems, just consider the fact that the
correct definition of a Hilbert space has been a very elusive
issue, i.e., a mathematically consistent definition of inner
product remains as a central problem in this model \cite{kopar04}.
In addition it has another complication, namely, the presence of
the concept of time. Indeed, along this very sophisticated
quantization method the concept of time has been lost
\cite{maca08}.

Though some advocates of all these models claim that they are on
the verge of a meaningful breakthrough, a reasonable doubt floats
upon these arguments. In connection with this last comment let us
add that there is a lack of experimental predictions in relation
with these approaches. In other words, they provide no possibility
for their testing. This current condition has spurred the search
for information which could provide some clue about the correct
direction. This topic is usually denoted as quantum gravity
phenomenology, i.e., the observational and experimental search for
deviations from Einstein's General Relativity and quantum theory.
It has to be stressed that nowadays it is a very active realm. In
this context string theory and loop gravity entail conjectures
which imply small modifications from General Relativity. For
instance, deviations from the $1/r$--potential and violations of
the equivalence principle \cite{acclamhm,FischbachTalmadge99} or
deformed versions of the dispersion relation \cite{Giovanni5}.

Both loop gravity \cite{AlfraoMoralesTecotlUrrutia00} and string
theory, \cite{Ellisetal99} imply induced modifications to the
field equations governing the motion of
spin--$\frac{1}{2}$--particles. Based upon string theory,
Kostelecky and coworkers have initiated an intensive study of
modifications of the standard model of elementary particle physics
\cite{CollodayKostelecky97, kos99, almourr02, ColemanGlashow97}.
In the search of experimental tests the main difficulty lies in
the smallness of the predicted effects. The proposed experiments
range from modifications to the standard model of elementary
particle physics \cite{Bluhm, Mattingly}, interferometric tests
\cite{Giovanni2, Giovanni3}, modification of Maxwell's equations
\cite{lamaho05}, or the use of Bose--Einstein condensates
\cite{Dalfovoro} in this context
\cite{Colladay,Donald,CastellanosCamacho, CastellanosCamacho1}. A
fruitful realm in the direction of precision tests is atomic
physics. Indeed, the Hughes--Drever experiment shows us the power
of this resource \cite{Will93}. One of the advantages of atomic
physics is related to the fact that some of most accurately tested
effects lie within this context. Indeed, the relative experimental
uncertainty related to the measurement of the hyperfine splitting
of the ground state of the hydrogen atom can be considered among
the most accurate experiments, i.e., it does not exceed $10^{-12}$
\cite{Dupays}.

The idea in the present work is to take advantage of this high
experimental precision for the fine and hyperfine effects and
determine bounds for some quantum gravity effects. In particular
we will address the issue of the consequences of the presence of
metric fluctuations \cite{Ertan1, Ertan2} upon the properties of a
hydrogen atom. The main ingredient in this aspect corresponds to a
Minkowskian background and in addition small spacetime
fluctuations are also present. One of the assumptions in this
approach is related to the fact that these spacetime fluctuations
emerge as classical fluctuations of the metric, in our case the
Minkowskian metric.

In addition, we will introduce these metric fluctuations in the
corresponding effective Schr\"odinger equation and deduce the
modifications that they entail upon the hyperfine structure
related to a hydrogen atom. We will find the change that these
effects imply for the ground state energy of the system. Once
again, we will set a bound to these oscillations resorting to the
current experimental outcomes \cite{Dupays}. Indeed, these
fluctuations can be comprehended (at least partially) as
redefinitions of the inertial mass \cite{Ertan1, Ertan2}, the new
inertial mass for the electron ($m^{eff}_e$) is given by

\begin{equation}
m^{eff}_e =m_e\Bigl(1+\gamma\Bigr)^{-1}.\label{elemass1}
\end{equation}

Here $m_e$ and $m_n$ are the electronic and nuclear masses,
respectively. It has to be clearly stated that this parameter
($\gamma$) depends upon the type of particle.

\begin{equation}
\mu  = \frac{m_em_n}{m_e+m_n}.\label{redmass1}
\end{equation}

To lowest order in the parameter $m_e/m_n$ we have that the new
reduced mass of the system ($\mu^{eff}$) shows the effects of the
metric fluctuations (here $\gamma$ is related to the electron).
The absence of the corresponding variable for the proton stems
from the fact that it appears as a higher--order term, and we keep
here the dominant contribution.

\begin{equation}
\mu^{eff}  = \mu\Bigl(1+\gamma\Bigr)^{-1}.\label{redmass2}
\end{equation}

This last remark allows us to estimate, roughly, the effects of
the metric perturbations, i.e., the functional dependence upon
$\gamma$. Indeed, a fleeting glimpse at the modifications caused
by the fine and hyperfine terms for a hydrogen atom, here we take
as an example the case of the $1s$--level, shows that the fine
contribution reads: $\Delta E_f = -\frac{1}{8}\mu c^2\alpha^4$,
whereas the hyperfine modification is given by: $\Delta E_h \sim
\frac{m^2_e}{m_p}c^2\alpha^4$. In this sense $\Delta E \sim
\frac{\mu c^2\alpha^2}{2}\gamma$ cannot be related to a correction
to the fine structure contributions, i.e., it goes like $\sim
\alpha^2$, and not $\sim \alpha^4$. Here $c$ is the speed of light
and $\alpha$ stands for the fine structure constant \cite{Cohen1}.

We also analyze the corresponding Lamb shift effect \cite{Scully1}
emerging from these fluctuations of spacetime and obtain a bound
for its size comparing our theoretical predictions against the
experimental uncertainty reported in the literature. The
possibility of the emergence of a Lamb type--like shift induced by
the fluctuations of the metric is the issue to be addressed as a
part of the present work. The absence of a quantum electrodynamic
theory predicts, for the Hydrogen atom an accidental degeneracy
between the $2S_{1/2}$ and $2P_{1/2}$ levels, i.e., they have the
same energy. The formulation of a quantum version of
electrodynamics breaks down this aforementioned accidental
degeneracy. Indeed, the effects of the fluctuations of the
electric and magnetic fields with the vacuum entails a
perturbation to the solutions stemming from the Coulombian
potential. The Lamb shift has been a cornerstone in the
development of several areas of Physics, among them, atomic
physics and quantum electrodynamics. The experimental uncertainty
in this realm offers one of the best scenarios for precision
tests. For the sake of completeness let us provide an explanation
of how these metric fluctuations can give rise to an effect
similar to the Lamb shift. The point here is that in the usual
model \cite{Scully1} the electron and proton positions are
shifted, due to the fluctuations to the electromagnetic field,
$r\rightarrow r +\delta r$. This last and simple comment explains
in a very intuitive manner why metric fluctuations shall also
impinge upon this aspect.
\bigskip
\bigskip

\section{Atomic Structure and spacetime fluctuations}
\bigskip

\subsection{Perturbation Procedure}
\bigskip

The first part of the present work addresses the issue of the
effects of the these metric fluctuations upon the hyperfine levels
of a hydrogen atom. The particular structure of our model reads

\begin{equation}
\hat{W} = \frac{1}{2\mu}\gamma^{ij}\hat{P}_i\hat{P}_j, ~i, j= x,
y, z. \label{Fluc1}
\end{equation}

In this last expression $\hat{P}_i$ denotes the momentum operator
and $\gamma^{ij}$ are the effects upon the effective Schr\"odinger
equation of the metric fluctuations \cite{Ertan1, Ertan2}, and
$\mu$ the reduced mass of the hydrogen atom.

This last operator $\hat{W}$ will be considered as a perturbation
upon the eigenkets associated to a hydrogen atom. In these
eigenkets the fine and hyperfine effects will be included, i.e.,
our initial eigenkets shall contain the information involving fine
and hyperfine structures. At this point we must explain the
reasons behind this procedure. Indeed, it is already known that
the fine structure Hamiltonian has a large magnitude than the one
related to the hyperfine effects \cite{Cohen1}, their ratio goes
like $\alpha^2$, i.e., the square of the fine structure constant.
This argument tells us that we must first calculate the
modifications upon energies and kets of the hydrogen atom
resorting to the fine structure Hamiltonian. Afterwards, the
hypefine terms will be introduced into the energies and kets
obtained in the first step. Finally, since we assume, from the
very beginning, that (\ref{Fluc1}) is a very tiny contribution to
the atomic behavior, i.e., the smallest of all of them, and, in
consequence, according to perturbation theory it has to be the
last term to be included in this approximation procedure.

The idea is to find a bound for the order of magnitude associated
to $\gamma^{ij}$ resorting to the comparison between our
theoretical predictions and the current experimental bounds. As
previously mentioned, the relative experimental uncertainty
related to the measurement of the hyperfine splitting of the
ground state of the hydrogen atom can be considered among the most
accurate experiments, i.e., it does not exceed $10^{-12}$
\cite{Dupays}.

Let us for a moment consider only the effects of (\ref{Fluc1}).
The most general case, concerning the structure of the spacetime
fluctuations, does not consider any kind of condition upon the
involved parameters, namely, $\gamma^{ij}$. Under this general
situation it turns out that some properties of the hydrogen atom,
stemming from spherical symmetry, will be lost. For instance,
(\ref{Fluc1}) entails the presence of three different spherical
tensor operators. Indeed, any vector operator, like the momentum
operator $(P_x, P_y, P_z)$, defines, uniquely, a spherical tensor
of rank $k=1$ \cite{Sakurai}, $T^{(k=1)}_{(q=0)} = P_z$, whereas,
$T^{(k=1)}_{(q=\pm 1)} =\frac{\mp 1}{\sqrt{2}}\Bigl\{P_x \pm
iP_y\Bigr\}$. An additional theorem \cite{Sakurai} implies that an
expression like (\ref{Fluc1}) contains three different types of
spherical tensors, namely, of ranks $k=0, 1, 2$ (a consequence of
the conditions that the Clebsch--Gordan coefficients satisfy).
These last arguments entail that, for instance, the spherical
tensor of rank $k=2$ associated to (\ref{Fluc1}) will break down
the inherent degeneracy of this atom. This can be understood
noting that if we take two different kets of the hydrogen atom,
with the same energy (same quantum number $n$) and same angular
momentum (same $l$) then according to Wigner--Eckardt theorem
\cite{Sakurai}.

\begin{equation}
<n, l, m\vert T^{(k=2)}_{(q)}\vert n, l,\hat{m}> =
\frac{<n,l\vert\vert T^{(k=2)}\vert\vert n,
l>}{\sqrt{2l+1}}<l,\hat{m};k=2, q\vert l, k=2; l, m>.
\label{Cleb1}
\end{equation}

The rules satisfied by the Clebsch--Gordan coefficients imply that
(\ref{Cleb1}) vanishes if $\hat{m} + q\not = m$ or if $l\not\in
[\vert k-l\vert, k+l]$ \cite{Sakurai}. In other words,

\begin{eqnarray}
<n=2, l=1, m=+1\vert T^{(k=2)}_{(q)}\vert n=2, l=1,\hat{m}=-1>
\not= \nonumber\\
<n=2, l=1, m=+1\vert T^{(k=2)}_{(q)}\vert n=2,
l=1,\hat{m}=+1> . \label{Cleb2}
\end{eqnarray}

According to perturbation theory of degenerate levels the last
expression leads us to conclude that (in the subspace of $l=1$)
the matrix defined by (\ref{Cleb1}) is not a matrix proportional
to the identity matrix, i.e., it has more than one eigenvalue and
eigenvector. These last arguments tell us, in a rough way, the
modifications due to these metric fluctuations. Notice that for
$l=0$, the case in which we are interested, one of the
Clebsch--Gordan conditions ($l\not\in [\vert k-l\vert, k+l]$)
means that only for $k=0$ it could be different from zero, because
$0 \not\in [2,2]$ and $0 \not\in [1,1]$. In other words, the
four--fold degeneracy of the ground state (taking into account
electronic and protonic spins) will not be broken by anisotropic
fluctuations. It is readily seen that this last comment entails
the fact that spherical symmetry cannot be broken. In general the
role of metric fluctuations becomes richer if the restriction of
conformal condition is released. In this case, spherical symmetry
will be broken and the degeneracy, inherent and accidental, can be
lost.

In the present work the simplest case will be considered, namely,
$\gamma^{ij} = 0$, if $i\not =j$, whereas $\gamma^{xx} =
\gamma^{yy} = \gamma^{zz} =\gamma$. This is due to the fact that
we are interested in the case $l=0$. Clearly, as already shown,
the situation can not lead to the breakdown of spherical symmetry,
though it does not mean that the case is uninteresting. Indeed, it
leads to a modification of the ground energy of the hydrogen atom,
and the modification of this energy due to the fluctuations has
the structure (this statement will be proved latter) $\Delta E
\sim \frac{\mu c^2\alpha^2}{2}\gamma$, in the leading
contribution.
\bigskip

\subsection{Hierarchy of perturbation terms}
\bigskip

The Fine Structure Hamiltonian reads \cite{Cohen1}

\begin{equation}
\hat{W}_{f} = -\frac{P^4}{8\mu^3c^2} +
\frac{1}{2\mu^2c^2}\frac{1}{r}\frac{dV}{dr}\vec{L}\cdot\vec{S} +
\frac{\hbar^2}{8\mu^2c^2}\nabla^2 V(r). \label{Fine1}
\end{equation}

Here $V(r)$ denotes the Coulombian potential. Let us, briefly,
explain each one of the three contributions appearing in the
right--hand side of this last expression. The first term can be
understood as the first relativistic correction to the dispersion
relation. The second one takes into account the fact that the
electron moves with respect to the nucleus and, therefore,
according to special relativity, it must {\it feel} the magnetic
field of the nucleus. Finally, the last term, called Darwin's
operator, is a consequence of the {\it Zitterbewegung} of the
electron \cite{Cohen1}.

There is an additional contribution that has to be included,
namely, the hyperfine \cite{Cohen1}.

\begin{equation}
\hat{W}_{h} = -\frac{\mu_0}{4\pi}\Bigl[\frac{q}{\mu
r^3}\vec{L}\cdot\vec{M}_I +
\frac{1}{r^3}\Bigl(3(\vec{n}\cdot\vec{M}_I)(\vec{n}\cdot\vec{M}_S)
- \vec{M}_S\cdot\vec{M}_I\Bigr) +
\frac{8\pi}{3}\vec{M}_S\cdot\vec{M}_I\delta(r)\Bigr].\label{Hfine1}
\end{equation}

In this last expression the first term on the right--hand side
describes the interaction of the nuclear magnetic moment with
magnetic field created by the orbital angular momentum of the
electron; the middle operator is related to the dipole--dipole
interaction between the magnetic moments of the electron and the
nucleus; whereas the last element, called Fermi's contact term, is
connected to the fact that the nucleus is not a point, i.e., it
has a non--vanishing spatial extension.

The procedure to be followed here is: (i) the initial eigenkets
will be those associated to the hydrogen atom \cite{Cohen2}. The
fine structure Hamiltonian will be used as a perturbation and the
corresponding perturbed energies and kets will be obtained. (ii)
Then these last results will be employed as original energies and
kets and the hyperfine structure Hamiltonian will be introduced as
a perturbation. (iii) Once again, the corresponding energies and
kets will be deduced and they will be the starting data and
(\ref{Fluc1}) will at this point play the role of a
time--independent perturbation. Clearly, this entails that the
effects of the fluctuations will be calculated in such a way that
fine and hyperfine effects are included as fundamental elements of
atomic physics.

We consider conformal fluctuations \cite{Ertan1, Ertan2}, namely,
$\gamma^{ij} = 0$, if $i\not =j$, whereas $\gamma^{xx} =
\gamma^{yy} = \gamma^{zz} = \gamma$. These conditions reduce
(\ref{Fluc1}) to

\begin{equation}
\hat{W} = \frac{1}{2\mu}\gamma\hat{P}^2. \label{Fluc2}
\end{equation}

At this point it seems to be a plain and simple re--definition of
the concept of inertial mass, i.e., $\mu\rightarrow
\frac{\mu}{\gamma}$, and we may wonder if this kind of effects can
be detected. The answer to this question is a very simple one and
can be understood recalling the general form of the fine structure
effect. Indeed, the energy modification due to this contribution
is given by $\Delta E_f\sim \frac{1}{8}\mu c^2\alpha^4$. If we
introduce the aforementioned redefinition we obtain that the new
modification reads $\Delta E_f\sim \frac{1}{8}\mu
c^2\alpha^4(1+\gamma)^{-1}$. Therefore, in principle, it can be
detected.

\bigskip

\subsection{Fine Structure}
\bigskip

\subsubsection{General Procedure}
\bigskip

We will focus on the modifications to the ground state energy, the
ionization energy. Since the fine structure effects are larger
than the hyperfine one,
$\vert\hat{W}_{h}/\hat{W}_{f}\vert\sim\alpha^2$ \cite{Cohen1},
then we start our analysis considering first fine structure terms
(\ref{Fine1}). The ground state of the hydrogen atom has a
four--fold degeneracy (due to the spin degrees of freedom of the
electron and proton). Since $l=0$, for the ground state, then the
spin--orbit coupling does nor contribute. In other words, the fine
structure cannot entail the breakdown of the spherical symmetry
for the ground state. This comment also implies that for $l=0$
$\hat{W}_{f}$ has a diagonal matrix. Hence the four corresponding
states move in the same way, namely, the first order correction to
these eigenstates reads

\begin{equation}
\vert 1> =
\sum_{n=2}^{\infty}\sum_{l=0}^{\infty}\sum_{m=-l}^{l}\frac{<n; l;
m; m_s; m_I\vert\hat{W}_{f}\vert n=1; l=0; m=0; m_s;
m_I>}{-\frac{1}{2}\mu c^2\alpha^2+ \frac{1}{2n^2}\mu
c^2\alpha^2}\vert n; l; m; m_s; m_I>. \label{kcorr1}
\end{equation}

There is an additional property which simplifies this last
expression. Indeed, the Darwin term vanishes for all states such
that $l\not =0$ \cite{Cohen1}.
\bigskip

\subsubsection{Matrix Elements}
\bigskip

Let us now provide an explanation that allows us to conclude that

\begin{equation}
\vert 1> =
\alpha^2\sum_{n=2}^{\infty}\frac{n^2}{n^2-1}\Bigl[\frac{1}{n^{3/2}}
-\frac{1}{n^3}\Bigr]\vert n; l=0; m=0; m_s; m_I>. \label{kcorr3}
\end{equation}

In this direction notice that $H_0 = \frac{P^2}{2\mu}
-\frac{q^2}{r}$, hence $\frac{P^4}{8\mu^3c^2} =\frac{1}{2\mu
c^2}(H_0 + \frac{q^2}{r})^2$, and in consequence we deduce

\begin{equation}
<n; l, m\vert \frac{P^4}{8\mu^3c^2}\vert \hat{n}; \hat{l},
\hat{m}> = \frac{1}{2\mu c^2}<n; l, m\vert (H_0 +
\frac{q^2}{r})^2\vert \hat{n}; \hat{l}, \hat{m}>. \label{kcorr13}
\end{equation}

From this last condition it turns out

\begin{eqnarray}
\frac{1}{2\mu c^2}<n; l, m\vert (H_0 + \frac{q^2}{r})^2\vert
\hat{n}; \hat{l}, \hat{m}> = \frac{1}{2\mu
c^2}\Bigl\{\Bigl(E_n\Bigr)^2\delta_{n,\hat{n}}\delta_{l,\hat{l}}\delta_{m,\hat{m}}
\nonumber\\
-2E_n<n; l, m\vert\frac{q^2}{r}\vert \hat{n}; \hat{l}, \hat{m}> +
<n; l, m\vert\frac{q^4}{r^2}\vert \hat{n}; \hat{l}, \hat{m}>
\Bigr\}. \label{kcorr14}
\end{eqnarray}

We now mention the fact that these two operators ($\frac{q^2}{r}$
and $\frac{q^4}{r^2}$) are both spherical operators of rank $k=0$,
i.e., they can not modify the orbital angular momentum or the
projection of it along the $z$--axis. This last remark entails

\begin{eqnarray}
<n; l, m\vert\frac{1}{r}\vert \hat{n}; \hat{l}, \hat{m}> =
\delta_{l,\hat{l}}\delta_{m,\hat{m}}\int_{0}^{\infty}R_{(n,
l)}(r)R_{(\hat{n}, \hat{l})}(r)rdr, \label{kcorr15}
\end{eqnarray}

\begin{eqnarray}
<n; l, m\vert\frac{1}{r^2}\vert \hat{n}; \hat{l}, \hat{m}> =
\delta_{l,\hat{l}}\delta_{m,\hat{m}}\int_{0}^{\infty}R_{(n,
l)}(r)R_{(\hat{n}, \hat{l})}(r)dr, \label{kcorr16}
\end{eqnarray}

Here the wavefunctions are $\psi_{(n,l,m)}= R_{(n,
l)}(r)Y_{(m)}^{(l)}$ \cite{Cohen2}.

\begin{eqnarray}
R_{(n, l)}(r)=
\Bigl[\Bigl(\frac{2}{na_0}\Bigr)^3\Bigl(\frac{(n-l-1)!}{2n(n+l)!}\Bigr)\Bigr]^{1/2}\exp\Bigl\{-\frac{r}{na_0}\Bigr\}
\Bigl(\frac{2r}{na_0}\Bigr)^l\mathcal{L}_{n-l-1}^{2l+1}\bigl(\frac{2r}{na_0}\bigr).
\label{kcorr17}
\end{eqnarray}

Here $\mathcal{L}_{n-l-1}^{2l+1}\bigl(\frac{2r}{na_0}\bigr)$
denotes the Laguerre associated polynomials, and $a_0$ the Bohr
radius \cite{Cohen2}.

Additionally, it is already known ($Re(\epsilon) >0$ and $s>0$)
\cite{Grad}

\begin{eqnarray}
\int_0^{\infty}e^{-\epsilon x}x^sdx = \frac{s!}{\epsilon^{s+1}}.
\label{Inte}
\end{eqnarray}

Resorting to this integral and the explicit expression for
Laguerre polynomials \cite{Grad} we may write down

\begin{eqnarray}
\int_{0}^{\infty}R_{(n, l)}(r)R_{(\hat{n},l)}(r)rdr =
\frac{1}{n\hat{n}a_0}\Bigl[\Bigl(\frac{(n-l-1)!}{(n+l)!}\Bigr)\Bigl(\frac{(\hat{n}-l-1)!}{(\hat{n}+l)!}\Bigr)\Bigr]^{1/2}
\times\nonumber\\
\sum_{s=0}^{\hat{n}-l-1}\sum_{t=0}^{n-l-1}\Bigl\{\frac{(-1)^{s+t}(n+l)!(\hat{n}+l)!(\hat{n}n)^{2l+3+s+t}}{t!s!(n-l-1-t)!(\hat{n}-l-1-s)!}\nonumber\\
\times\frac{(2l+s+t+1)!}{(s+2l+1)!(t+2l+1)!}
\frac{2^{2l+2+s+t}}{(n)^{l+t}(\hat{n})^{l+s}(n+\hat{n})^{2l+s+t+2}}\Bigr\}.
\label{kcorr18}
\end{eqnarray}

In order to calculate our required expressions let us point out
that since $H_0 = \frac{P^2}{2\mu} - \frac{q^2}{r}$ we have that

\begin{eqnarray}
<n; l, m\vert\frac{P^2}{2\mu}\vert \hat{n}; \hat{l},\hat{m}>= -\mu
c^2\alpha^2\Bigl[\frac{1}{2n^2}\delta_{n,\hat{n}} - <n; l,
m\vert\frac{q^2}{r}\vert \hat{n}; \hat{l},\hat{m}>\Bigr].
\label{kcorr19}
\end{eqnarray}

We mention this expression because it will help us out in the
evaluation of (\ref{kcorr18}). Indeed, the value of this last
expression when $n = \hat{n}$ is already known \cite{Robb},

\begin{eqnarray}
<n; l, m\vert\frac{P^2}{2\mu}\vert n;l,m>= \frac{\mu
c^2\alpha^2}{2n^2}. \label{kcorr20}
\end{eqnarray}

Therefore we conclude that if in (\ref{kcorr18}) we impose the
condition $n=\hat{n}$ (here we drop out the term $1/a_0$ since it
plays no role in the evaluation of this summation)

\begin{eqnarray}
\frac{1}{n^2} =
\frac{1}{n^2}\Bigl[\Bigl(\frac{(n-l-1)!}{(n+l)!}\Bigr)\Bigl(\frac{(n-l-1)!}{(n+l)!}\Bigr)\Bigr]^{1/2}
\times\nonumber\\
\sum_{s=0}^{n-l-1}\sum_{t=0}^{n-l-1}\Bigl\{\frac{(-1)^{s+t}(n+l)!(n+l)!(n)^{4l+6+2s+2t}}{t!s!(n-l-1-t)!(n-l-1-s)!}\nonumber\\
\times\frac{(2l+s+t+1)!}{(s+2l+1)!(t+2l+1)!}
\frac{2^{2l+2+s+t}}{(n)^{l+t}(n)^{l+s}(2n)^{2l+s+t+2}}\Bigr\}.
\label{kcorr21}
\end{eqnarray}

Immediately we see that

\begin{eqnarray}
1=
\Bigl[\Bigl(\frac{(n-l-1)!}{(n+l)!}\Bigr)\Bigl(\frac{(n-l-1)!}{(n+l)!}\Bigr)\Bigr]^{1/2}
\times\nonumber\\
\sum_{s=0}^{n-l-1}\sum_{t=0}^{n-l-1}\Bigl\{\frac{(-1)^{s+t}(n+l)!(n+l)!(n)^{4l+6+2s+2t}}{t!s!(n-l-1-t)!(n-l-1-s)!}\nonumber\\
\times\frac{(2l+s+t+1)!}{(s+2l+1)!(t+2l+1)!}
\frac{2^{2l+2+s+t}}{(n)^{l+t}(n)^{l+s}(2n)^{2l+s+t+2}}\Bigr\}.
\label{kcorr23}
\end{eqnarray}
\bigskip

Let us, at this point, sum up our conclusions concerning this
summation. On one hand, we know the result if $n= \hat{n}$. On the
other hand, we point out the fact that in (\ref{kcorr18}) the
summations in $s$ and $t$ are, functionally, the same; they differ
only in the upper limit, i.e., it is either $n-l-1$, or
$\hat{n}-l-1$. Let us now delve a little bit deeper in this
direction. In order to do this let us define

\begin{eqnarray}
g(s,t;n,\hat{n}) =
\Bigl[\Bigl(\frac{(n-l-1)!}{(n+l)!}\Bigr)\Bigl(\frac{(\hat{n}-l-1)!}{(\hat{n}+l)!}\Bigr)\Bigr]^{1/2}
\times\nonumber\\
\sum_{s=0}^{\hat{n}-l-1}\sum_{t=0}^{n-l-1}\Bigl\{\frac{(-1)^{s+t}(n+l)!(\hat{n}+l)!(\hat{n}n)^{2l+3+s+t}}{t!s!(n-l-1-t)!(\hat{n}-l-1-s)!}\nonumber\\
\times\frac{(2l+s+t+1)!}{(s+2l+1)!(t+2l+1)!}
\frac{2^{2l+2+s+t}}{(n)^{l+t}(\hat{n})^{l+s}(n+\hat{n})^{2l+s+t+2}}\Bigr\}.\label{kcorr24}
\end{eqnarray}

This remark concerning the functional dependence upon $s$ and $t$
means that if

\begin{eqnarray}
\sum_{t=0}^{n-l-1}g(s,t; n, \hat{n}) = f(s, n-l; n,\hat{n})
,\label{kcorr34}
\end{eqnarray}

 then

\begin{eqnarray}
\sum_{s=0}^{\hat{n}-l-1}g(s,t; n, \hat{n}) =
f(\hat{n}-l,t;n,\hat{n}). \label{kcorr35}
\end{eqnarray}

Where $f$ is the same function in both equations. Of course, we
have, on the right--hand side of (\ref{kcorr34}) and
(\ref{kcorr35}), the same function as a consequence of this
aforementioned equality in the functional dependence in terms of
$s$ and $t$. Clearly, this fact implies that

\begin{eqnarray}
\sum_{s=0}^{\hat{n}-l-1}\sum_{t=0}^{n-l-1}g(s,t; n, \hat{n}) =
h(n-l, \hat{n}-l),\label{kcorr36}
\end{eqnarray}

\noindent is a function of $n-l$ and $\hat{n}-l$ in which the
functional dependence upon these two variables is the same. In
addition, $h(n-l, n-l) = 1$, and this $\forall n\in \mathbb{N}$,
here $l$ is a constant parameter.

We now proceed to prove that $h(n, \hat{n}) = j(n)j(\hat{n})$,
i.e., we have $l=0$. Indeed, notice that our task is the
evaluation of the integral shown in (\ref{kcorr18}). In the
mathematical literature there are several variants denoted the
second mean value theorem for integrals. The usual ones cannot be
used in the present context since they require as a premise that
(at least) one of the two functions under the sign of integral has
to be an integrable positive function \cite{Burkhill}. In the
present case since $R_{(n, l)}(r)$ involves the Laguerre
associated polynomials then they may change its sign, and this
fact violates the premise of this theorem. The correct theorem is
a generalization of the second mean value theorem for integrals
done by Okamura \cite{Okamura}; if $f: [a, b]\rightarrow
\mathbb{R}$ is a monotonically decreasing function and $g: [a,
b]\rightarrow \mathbb{R}$ an integrable function, then there is
$c\in(a, b]$ such that

\begin{eqnarray}
\int_a^b f(x)g(x)dx = f(c)\int_a^b g(x)dx.\label{menin1}
\end{eqnarray}

It is already known that for the case of $l\not=0$ the
wavefunction for the hydrogen atom vanishes at the origin, this
fact is a consequence of the continuity of the solutions to this
potential \cite{Cohen1, Cohen2}. This remark entails that for
$l\not=0$ the radial part of the solution is an increasing
function from $r=0$ to a certain value, say $r_0$. This is not the
case for those radial parts of the solution associated to $l=0$,
since in this case they do not vanish at $r=0$ \cite{Cohen1,
Cohen2}. They have a decreasing behavior from $r=0$ to a certain
value given by the corresponding associated Laguerre polynomial.
Indeed, if $l=0$ we have that

\begin{eqnarray}
R_{(n, l=0)}(r)=
\Bigl[\Bigl(\frac{2}{na_0}\Bigr)^3\Bigl(\frac{1}{2n^2}\Bigr)\Bigr]^{1/2}\exp\Bigl\{-\frac{r}{na_0}\Bigr\}
\Bigl(1 - \bigl[n-1\bigr]\bigl\{\frac{r}{na_0}\bigr\}
\nonumber\\
+ \bigl[n-1\bigr]\bigl\{\frac{r}{na_0}\bigr\}^2
+...+(-1)^{n-1}\bigl[n-1\bigr]\bigl\{\frac{r}{na_0}\bigr\}^{n-1}\Bigr).
\label{Lag11}
\end{eqnarray}

This last expression shows us that between for $r\in[0,
n(n-1)a_0], \forall n\geq 2$, this is a decreasing function. In
addition, we know that for large values of $r$ the function is
clearly monotonically decreasing, i.e., they represent bound
states. These arguments allow us to accept (as a good
approximation) that there is $c\in(0, \infty)$

\begin{eqnarray}
\int_{0}^{\infty}R_{(n, l=0)}(r)R_{(\hat{n},l=0)}(r)rdr = R_{(n,
l=0)}(c)\int_{0}^{\infty}R_{(\hat{n},l=0)}(r)rdr . \label{menin2}
\end{eqnarray}

This last expression is the product of a function depending only
upon $n$ ($R_{(n, l=0)}(c))$ and another whose independent
variable is $\hat{n}$ ($\int_{0}^{b}R_{(\hat{n},l=0)}(r)rdr $).

This argument proves that $h(n, \hat{n}) = j(n)j(\hat{n})$, i.e.,
our function $h(n, \hat{n})$ can be written as the product of two
functions, one depending on $n$ and the other upon $\hat{n}$.

 Joining these two features we have $h(n, n)= [j(n)]^2 =1 \Rightarrow j(n) =\pm
 1$. This leads us to conclude that

\begin{eqnarray}
\pm
1=\Bigl[\Bigl(\frac{(n-1)!}{(n)!}\Bigr)\Bigl(\frac{(\hat{n}-1)!}{(\hat{n})!}\Bigr)\Bigr]^{1/2}
\times\nonumber\\
\sum_{s=0}^{\hat{n}-1}\sum_{t=0}^{n-1}\Bigl\{\frac{(-1)^{s+t}(n)!(\hat{n})!(\hat{n}n)^{3+s+t}}{t!s!(n-1-t)!(\hat{n}-1-s)!}\nonumber\\
\times\frac{(s+t+1)!}{(s+1)!(t+1)!}
\frac{2^{2+s+t}}{(n)^{t}(\hat{n})^{s}(n+\hat{n})^{s+t+2}}\Bigr\},\label{kcorr40}
\end{eqnarray}

The ambiguity concerning the sign appearing in this last
expression fades away noting that the integral (\ref{kcorr18}) has
to be non--negative, i.e., we must take the plus sign.

We may now write down
\begin{eqnarray}
\int_{0}^{\infty}R_{(n, l=0)}(r)R_{(\hat{n},l=0)}(r)rdr =
\frac{1}{n\hat{n}a_0}. \label{kcorr41}
\end{eqnarray}

The same kind of arguments can be used if the operator $1/r$ is
replaced by $1/r^2$. These results are to be introduced in the
corresponding matrix elements of (\ref{kcorr1}) and, in this way,
we are led to (\ref{kcorr3}) which corresponds to the particular
case in which $\hat{n} = 1$.

We may now, with these results, proceed to the calculation of the
perturbed kets and energies.

\subsubsection{Perturbed kets and energies}
\bigskip

The kets related to the new ground state read

\begin{eqnarray}
\vert\tilde{0}> = \vert n=1; l=0; m=0; m_s; m_I> \nonumber\\
+\alpha^2\sum_{n=2}^{\infty}\frac{n^2}{n^2-1}\Bigl[\frac{1}{n^{3/2}}
-\frac{1}{n^3}\Bigr]\vert n; l=0; m=0; m_s; m_I>. \label{kcorr4}
\end{eqnarray}
\bigskip

Notice that this last expression tells us that the dependence of
the perturbed ket upon $\vert n; l=0; m=0; m_s; m_I>$, with $n\geq
2$ is much smaller than that related to the case $n=1$. Indeed, it
is readily seen that those kets associated to $n\geq 2$ are
multiplied by $\alpha^2 \sim 10^{-4}$. This assertion is a
consequence of the fact that the fine structure terms behave as
$\alpha^4$, and when the corresponding matrix element is divided
by the difference between the two involved states ($\Delta E\sim
\frac{\mu c^2\alpha^2}{2}$) we obtain this aforementioned
behavior.

\subsection{Hyperfine Structure}
\bigskip

This new ground state ($\vert\tilde{0}>$) will be now employed for
the deduction of the first--order energy correction due to the
hyperfine structure terms (\ref{Hfine1}).

\begin{equation}
E_1 = <\tilde{0}\vert\hat{W}_{h}\vert\tilde{0}>.\label{kcorr5}
\end{equation}

Explicitly, this energy becomes

\begin{eqnarray}
E_1 = -g_p\frac{\mu_e}{m_p}\bigl(\mu_ec^2\alpha^4\bigr)\Bigl\{1 +
\sum_{n=2}^{\infty}\frac{n^2\alpha^2}{n^2-1}\Bigl[-\frac{1}{n^{9/2}}+ \frac{1}{n^{3}}\Bigr]\nonumber\\
+ \Bigl(\sum_{n=2}^{\infty}\frac{n^2\alpha^2}{n^2-1}\Bigl[
-\frac{1}{n^{9/2}}+
\frac{1}{n^{3}}\Bigr]\Bigr)^2\Bigr\}.\label{kcorr6}
\end{eqnarray}

Riemann's function ($\xi(\nu) = \sum_{l=1}^{\infty}1/l^{\nu}$)
allows us to find a simple expression to the first order energy
correction. Indeed, our approximation will be

\begin{equation}
\sum_{n=2}^{\infty}\frac{n^2}{n^s(n^2-1)}\approx
\frac{4}{3}\Bigl[\xi(s) -1\Bigr]. \label{reimann3}
\end{equation}

Let us explain this assumption. Notice that

\begin{equation}
\sum_{n=2}^{\infty}\frac{n^2}{n^s(n^2-1)}=
\sum_{n=2}^{\infty}\frac{n^2-1+1}{n^s(n^2-1)} =
\sum_{n=2}^{\infty}\frac{1}{n^s} +
 \sum_{n=2}^{\infty}\frac{1}{n^s(n^2-1)}, \label{error1}
\end{equation}

\begin{equation}
\sum_{n=2}^{\infty}\frac{n^2}{n^s(n^2-1)}= \xi(s)-1 +
\sum_{n=2}^{\infty}\frac{1}{n^s(n^2-1)}. \label{error2}
\end{equation}

We now take the second term on the right--hand side of this last
expression

\begin{equation}
\sum_{n=2}^{\infty}\frac{1}{n^s(n^2-1)} = \frac{1}{3\cdot 2^s} +
\frac{1}{8\cdot 3^s}...\leq
\frac{1}{3}\sum_{n=2}^{\infty}\frac{1}{n^s} =
\frac{1}{3}\Bigl(\xi(s)-1\Bigr). \label{error3}
\end{equation}

Hence

\begin{equation}
\sum_{n=2}^{\infty}\frac{n^2}{n^s(n^2-1)}\leq
\frac{4}{3}\Bigl(\xi(s)-1\Bigr). \label{error4}
\end{equation}

In order to have a deeper understanding of this approximation we
have calculated some {\it exact} results as well as our
expression. The relative error between these two cases has also
been carried out. This has been done for three different values of
$s$, as the table below shows. The choice for these values of $s$
has been done in terms of their relevance for the final energy
expression. It is readily seen that the relative error decreases
as $s$ grows, it goes from, approximately, $10$ percent to $2$
percent.
\bigskip
\bigskip

\begin{center}
\begin{tabular}{|l|l|l|l|}
\hline \textbf{Parameter $s$} &
$\sum_{n=2}^{\infty}\frac{n^2}{n^s(n^2-1)}$&
$\frac{4}{3}\left(\zeta(s)-1\right)$ & rel. error\\\hline\hline
$s=5/2$&  0.41199220 &0.4553164 & 0.10515782\\\hline
$s=4$&0.10506593 & 0.1097643 & 0.044718588 \\\hline
$s=5$&0.047943097  & 0.0492371  & 0.026990330 \\\hline

\end{tabular}
\end{center}
\bigskip
\bigskip

These last arguments allow us to write the first--order energy
correction:

\begin{eqnarray}
E_1 = -g_p\frac{\mu_e}{m_p}\bigl(\mu_ec^2\alpha^4\bigr)\Bigl\{1 +
\frac{4\alpha^2}{3}\Bigl[-\xi(9/2)+\xi(3)\Bigr]\Bigr\}.\label{kcorr7}
\end{eqnarray}

The calculation of the perturbed kets has to be done resorting to
(\ref{kcorr4}). Once again, this ground state is four--fold
degenerate, nevertheless, let us mention that the term involving
$\vec{L}\cdot\vec{M}_I$ vanishes since our ground state implies
$l=0$. Similarly, the contribution involving the dipole--dipole
interaction turns out to be zero (due to spherical symmetry)
\cite{Cohen1}. The only term that participates is the last one,
the so--called contact term. Here we must calculate terms with the
following structure

\begin{eqnarray}
<n; l=0; m=0; m_s;
m_I\vert\frac{8\pi}{3}\vec{M}_S\cdot\vec{M}_I\delta(r)\vert
\tilde{n}; l=0; m=0; \tilde{m_s}; \tilde{m_I}> =
\nonumber\\
\frac{8\pi}{3}R_{(n, l=0)}(r=0)R_{(\tilde{n}(r=0),
l=0)}<m_s;m_I\vert\vec{M}_S\cdot\vec{M}_I\vert\tilde{m_s};
\tilde{m_I}>.\label{cont1}
\end{eqnarray}

In this last expression $R_{(n, l=0)}(r)$ denotes the radial part
of the corresponding wavefunction. We now define (as usual, since
$\vec{M}_S$ and $\vec{M}_I$ are proportional to $\vec{S}$ and
$\vec{I}$, respectively)

\begin{eqnarray}
\vec{F} = \vec{S} + \vec{I}.\label{cont2}
\end{eqnarray}

Therefore we have that

\begin{eqnarray}
\vec{F}^2 = \vec{S}^2 + \vec{I}^2 +
2\vec{I}\cdot\vec{S}.\label{cont3}
\end{eqnarray}

Here, according to the rules of addition of angular momentum,
$F=0, 1$. In the eigenkets of $\vec{F}^2$ (here denoted $\vert F,
m_F>$).

\begin{eqnarray}
 <n; F=0,m_F=0\vert\hat{W}_{h}\vert \hat{n}=1;F=0, m_F=0> =
 -\frac{q^2\hbar^2g_p\mu_0}{4\pi a_0^3n^{3/2}\mu_em_p},\label{cont4}
\end{eqnarray}

\begin{eqnarray}
 <n;F=1,m_F\vert\hat{W}_{h}\vert \hat{n}=1;F=1, m_F> =
 \frac{q^2\hbar^2g_p\mu_0}{12\pi a_0^3n^{3/2}\mu_em_p}.\label{cont5}
\end{eqnarray}

We may now write down the perturbed kets, which contain the
effects of fine and hyperfine structures. Notice that the
correction related to $F=0$ implies a lower energy than those
three cases associated to $F=1$. In other words, the ground state
now is given by

\begin{eqnarray}
\vert\hat{0}> = \vert n=1,; l=0, m_l=0; F=0, m_F=0> +\nonumber\\
+\alpha^2\sum_{n=2}^{\infty}\frac{n^2}{n^2-1}\Bigl[\frac{1}{n^{3/2}}
-\frac{1}{n^3}\Bigr]\vert n; l=0, m_l=0; F=0,
m_F=0>\nonumber\\
+
2\frac{g_p\mu_e}{m_p}\alpha^2\sum_{n=2}^{\infty}\frac{n^2}{n^2-1}\frac{1}{n^{3/2}}\vert
n; l=0, m_l=0; F=0, m_F=0> +...\label{cont6}
\end{eqnarray}

Now the consequences of the metric fluctuations upon this ground
state are deduced

\begin{eqnarray}
<\hat{0}\vert \frac{1}{2\mu}\gamma\hat{P}^2\vert\hat{0}> =
-\frac{1}{2}\mu c^2\alpha^2\gamma\Bigl\{1 -
\frac{16}{3}\alpha^2\Bigl[\xi(5/2)- \xi(4)+
2\frac{g_p\mu_e}{m_p}\bigl(\xi(5/2)-1\bigr)\Bigr] +
O\Bigl(\alpha^4\Bigr)\Bigr\}.\label{kcorr8}
\end{eqnarray}

Finally, the energy of the ground state, which is the ionization
energy is

\begin{eqnarray}
E_0 = -\frac{1}{2}\mu_e c^2\alpha^2\Bigl\{1 + \frac{1}{4}\alpha^2
+ 2g_p\frac{\mu_e}{m_p}\alpha^2\Bigl(1-1.82\gamma\Bigr)
+\gamma\bigl(1 - 1.376\alpha^2\bigr)\Bigr\} .\label{kcorr9}
\end{eqnarray}

If $\gamma\rightarrow 0$, then we recover the usual result
\cite{Cohen1}. It is also readily seen that the effects of metric
fluctuations appear here as a modification to the ionization
energy of the hydrogen atom

\begin{eqnarray}
\Delta E_0 = -\frac{1}{2}\mu_e c^2\alpha^2\gamma\Bigl\{1 -
1.376\alpha^2-3.64g_p\frac{\mu_e}{m_p}\alpha^2\Bigr\}
.\label{Unce1}
\end{eqnarray}

\section{Spacetime fluctuations and Lamb Shift}
\bigskip

We now address the issue of a possible connection between metric
fluctuations and a Lamb--type like shift \cite{Welton}. In
addition to the modification imposed by the zero--point
fluctuations of the electromagnetic field, since the potential
$V(r)$ is a function of $r$ then the presence of these spacetime
fluctuations entails a change of the position of the electron (of
course, also of the proton), $r\rightarrow r +\delta r$.

\begin{eqnarray}
\Delta V = V(r +\delta r) - V(r) = \delta\vec{r}\cdot\nabla V(r) +
\frac{1}{2}\Bigl(\delta\vec{r}\cdot\nabla\Bigr)^2V(r)+....\label{Lamb1}
\end{eqnarray}

The method provides the correct order of magnitude of this effect
since it reproduces the Bethe formula \cite{Welton,Bethe}. The
assumption of conformal fluctuations implies \cite{Ertan1, Ertan2}
that the average of this change (here denoted by $<\Delta V>$)

\begin{eqnarray}
<\Delta V> =
<\frac{1}{2}\Bigl(\delta\vec{r}\cdot\nabla\Bigr)^2V(r)>.
\label{Lamb2}
\end{eqnarray}

The meaning of $<\Delta V>$ entails two different averages: (i) an
average over metric fluctuations and (ii) a second average over
atomic states. Indeed, isotropy leads us to conclude that the
average over metric fluctuations

\begin{eqnarray}
<\delta\vec{r}>_{(mf)}=0 \label{Metric0}
\end{eqnarray}

Let us now explain how the average over metric fluctuations is to
be carried out. Here the background metric is the Minkowskian one,

\begin{eqnarray}
ds^2 = e^{\psi(x)}\eta_{00}dt^2 + e^{\psi(x)}\eta_{ij}dx^idx^j,
\label{Metric1}
\end{eqnarray}
\bigskip

We assume (with the condition $\vert\psi(x)\vert<<1$)

\begin{eqnarray} <\psi(x)>_{(mf)} = 0,\label{Metric3}
\end{eqnarray}

\begin{eqnarray} <\partial_{\mu}\psi(x)>_{(mf)} = 0.\label{Metric33}
\end{eqnarray}

Therefore we have

\begin{eqnarray} <\delta x> =
<e^{\psi(x)}x\eta_{xx}>_{(mf)} = \Bigl(1 +
\frac{1}{2}<(\psi)^2>_{(mf)}+...\Bigr)x\eta_{xx}, \label{Lamb2}
\end{eqnarray}

\begin{eqnarray} <e^{\psi(x)}\eta_{00}>_{(mf)} = \Bigl(1 +
\frac{1}{2}<(\psi)^2>_{(mf)}+...\Bigr)\eta_{00}. \label{Metric2}
\end{eqnarray}

\noindent and similarly for the remaining coordinates. The imposed
condition (\ref{Metric3}) implies (\ref{Metric0}). It is easily
seen that these conditions contain isotropy and homogeneity of
spacetime. Indeed, it does not define a privileged direction nor a
privileged point.

We now notice that, explicitly,

\begin{eqnarray}
\frac{1}{2}\Bigl(\delta\vec{r}\cdot\nabla\Bigr)^2V(r) =
\frac{\delta\vec{r}}{2}\cdot\Bigl\{\bigl(\delta\vec{r}\cdot\nabla\bigr)\nabla
V + \nonumber\\\bigl(\nabla V\cdot\nabla\bigr)\delta\vec{r} +
\delta\vec{r}\times\nabla\times\bigl(\nabla V\bigr) + \nabla
V\times\nabla\times\delta\vec{r}\Bigr\}. \label{Lamb3}
\end{eqnarray}

The conformal condition imposed upon these fluctuations allow us
to simplify the calculations. Indeed, isotropy and independence of
the fluctuations along the different directions can be rephrased
as follows

\begin{eqnarray}
<\delta x\delta y>_{(mf)} = <\delta x>_{(mf)}<\delta y>_{(mf)},
\label{Lamb3}
\end{eqnarray}

\begin{eqnarray}
<(\delta x)^2>_{(mf)} = \frac{1}{3}<(\delta\vec{r})^2>_{(mf)}.
\label{Lamb4}
\end{eqnarray}

Hence the average over the metric fluctuations can be reduced to

\begin{eqnarray}
\frac{1}{2}<\Bigl(\delta\vec{r}\cdot\nabla \Bigr)^2>_{(mf)} =
\frac{1}{6}<(\delta\vec{r})^2>_{(mf)}\nabla^2.\label{Lamb5}
\end{eqnarray}

 Up to now we have only introduced, explicitly,
metric fluctuations ($<>_{mf}$). At this point we require an
assumption for one of our parameters, namely, the behavior of some
of the statistical properties of the fluctuations have to be
given. We introduce \cite{Camacho}

\begin{eqnarray}
<(\delta\vec{r})^2>_{(mf)} =\sigma^2.\label{Lamb6}
\end{eqnarray}

In this last expression $\sigma$ is a constant. We  may now
conclude

\begin{eqnarray}
<\Delta V> = \frac{\sigma^2}{6}<\nabla^2(\frac{-q^2}{r})>_{(at)}
.\label{Lamb7}
\end{eqnarray}

This average ($<>_{(at)}$) denotes a calculation over atomic
states, i.e.,

\begin{eqnarray}
\frac{\sigma^2}{6}<\nabla^2(\frac{-q^2}{r})>_{(at)} =
-\frac{q^2\sigma^2}{6}\int\vert\psi_{(n,l,m)}(\vec{r})\vert^2\nabla^2\bigl(\frac{1}{r}\bigr)d^3r
.\label{Lamb8}
\end{eqnarray}

Finally, $\nabla^2\bigl(\frac{1}{r}\bigr)= -4\pi\delta(r)$, and we
obtain

\begin{eqnarray}
<\Delta V> =4\pi\frac{q^2\sigma^2}{6}\int\delta(r)
\vert\psi_{(n,l,m)}(\vec{r})\vert^2d^3r.\label{Lamb9}
\end{eqnarray}

This last result, as in the usual Lamb shift effect
\cite{Scully1}, implies that metric fluctuations, in this context,
can modify only those states with vanishing angular momentum
($l=0$), due to the fact that $\psi_{(n,l,m)}(\vec{0}) =0$ if
$l\not =0$ \cite{Cohen1}. Then, for the particular case of $n=2$
we have (here $a_0$ denotes Bohr radius)

\begin{eqnarray}
\frac{q^2\sigma^2}{6}\vert\psi_{(2,0,0)}(\vec{0})\vert^2 =
-\frac{1}{96}\Bigl(\frac{\sigma}{a_0}\Bigr)^2\frac{q^2}{a_0}.\label{Lamb10}
\end{eqnarray}

\bigskip
\bigskip

\section{Discussion and results}
\bigskip

Let us now discuss the results and their perspectives, always in
the context of precision tests.
\bigskip

\subsection{Hyperfine splitting}
\bigskip

Our result concerning the ionization energy of the hydrogen atom
is provided by (\ref{Unce1}).  The first term ($-(0.5)\mu_e
c^2\alpha^2\gamma$) can be comprehended as a consequence of the
re--definition of the concept of inertial mass, see
(\ref{elemass1}). Indeed, it is quadratic in the fine structure
constant, a fact that discards any relation with the fine or
hyperfine structure.

The second contribution ($(0.688)\mu_e c^2\alpha^4\gamma$) has to
be related to the fine structure. Of course, the dominant effect
(since $\alpha\sim 1/137$) corresponds to $-(0.5)\mu_e
c^2\alpha^2\gamma$. The emergence of $\gamma$ in this last term is
also a consequence of this re--definition of the inertial mass.
Indeed, take, for instance, the Darwin term,
$\frac{\hbar^2}{8\mu_e^2c^2}\nabla V(r)$. Its average for an
unperturbed eigenstate (with quantum numbers $n, l=0,m=0$) of the
hydrogen atom behaves like $\hbar^2/(8n^2\mu_e^2c^2a_0^3)$. The
fine structure constant is defined as $\alpha =\hbar/(\mu_e
ca_0)$, and therefore, this average has the order of magnitude
$\mu_e c^2\alpha^{4}/n^2$. If we re--define $\mu^{eff}
=\mu_e\Bigl(1+\gamma\Bigr)^{-1}$, then we deduce that the Darwin
term is also modified, namely, $\mu_e
c^2\alpha^{4}/n^2\rightarrow\mu_e
c^2\alpha^{4}\Bigl(1+\gamma\Bigr)^{-1}/n^2$. In other words, this
redefinition of the inertial mass is the first modification that
emerges in the present context, either in connection with the
unperturbed energy levels, or in association with the fine
structure terms.

The third and last contribution ($1.82\mu_e c^2\alpha^4\gamma
g_p\frac{\mu_e}{m_p}$) corresponds to the modifications upon the
hyperfine energy due to these fluctuations and, once again, it can
be understood as a consequence of this redefinition of the
inertial mass. The hyperfine splitting of the ground state of the
hydrogen atom can be considered among the most accurately detected
physical quantities \cite{Dupays}. The related uncertainty is
associated to several possibilities, for instance, correction
terms of higher--order QED effects, or the proton electromagnetic
structure induced by strong interactions. A very rough bound for
the magnitude of $\gamma$ can be deduced from \cite{Dupays},
equation (1). Indeed, according to this experimental result the
uncertainty related to the hyperfine splitting of the hydrogen
atom reads

\begin{eqnarray}
\Delta E^{(exp)}_0 = \pm 0.0009\label{Lamb66}~Hz.
\end{eqnarray}

We now relate our third term ($1.82\mu_e c^2\alpha^4\gamma
g_p\frac{\mu_e}{m_p}$) to the experimental uncertainty

\begin{eqnarray}
1.82\mu_e c^2\alpha^4\gamma g_p\frac{\mu_e}{m_p}\leq\hbar\Delta
E^{(exp)}_0\label{Lamb68}.
\end{eqnarray}

In this way we obtain a rough bound for $\gamma$

\begin{eqnarray}
\gamma \leq 10^{-18} .\label{Unce5}
\end{eqnarray}
\bigskip

\subsection{Lamb shift}
\bigskip

The Lamb shift has been used as a tool for the detection of the
radius of the proton, etc. \cite{Pohl}. In our case it allows us
to pose, in an independent way from the previous situation, a
bound to some statistical properties of the metric fluctuations.
Indeed, we may use the uncertainty ($\Delta E$) mentioned in
\cite{Hildum} ($\Delta E/\hbar = 4.8$MHz).

\begin{eqnarray}
\frac{1}{96}\Bigl(\frac{\sigma}{a_0}\Bigr)^2\frac{q^2}{a_0}
\leq\Delta E, \label{Lamb11}
\end{eqnarray}

\noindent and obtain as a bound
\begin{eqnarray} \sigma\leq
10^{-18}m.\label{Lamb12}
\end{eqnarray}
\bigskip
\bigskip
\bigskip

\subsection{Fluctuations Models and Precision Tests}
\bigskip

It has to be emphasized that these two bounds ((\ref{Unce5}) and
(\ref{Lamb12})) cannot be compared directly. Indeed, $\gamma$ is
not equal to $\sigma$ ($\gamma^{ij}$ in the general situation)
\cite{Ertan1, Ertan2}. They correspond to different statistical
features of the involved fluctuations of spacetime. Of course,
they are related and at this point we mention their connection.
The effective Schr\"odinger equation used in the context of
hyperfine structure contains an average process of metric
fluctuations involving a spacetime interval (see expression (3) in
\cite{Ertan1}). The averaging method involved in the calculations
of the Lamb shift \cite{Camacho} has a resemblance of the first
procedure, though it is not the same kind of fluctuation, as
expression (4) in \cite{Ertan1} explicitly manifests. Indeed, in
this last approach the average of these fluctuations defines a
background field, whereas in \cite{Camacho} the average of these
fluctuations does not define this parameter. This lack of
coincidence in the context of our employed fluctuations shall be
no surprise \cite{Crowell}, i.e., the possibilities in the type of
fluctuations is not restricted to those mentioned in \cite{Ertan1}
or in \cite{Camacho}. This fact can be considered a drawback of
quantum gravity phenomenology since a very large number of options
are to be included, as possible cases.

Concerning (\ref{Unce5}) our parameter $\gamma$ is closely related
(in a weak field approach) to perturbations of the Minkowskian
metric, in a similar way as in the case of linear gravitational
waves. It can be seen that the present bound is three orders of
magnitude larger than the strength of a gravitational wave whose
source is a stellar binary \cite{Schutz}.

In relation with (\ref{Lamb12}) we may also find a bound for
$\vert<\delta r>\vert$. Indeed, we know that the standard
deviation is defined as $\Delta r = \sqrt{\sigma^2 - (<\delta
r>)^2}$. Therefore we find (since $\Delta r \geq 0$) that
$\vert<\delta r>\vert\leq 10^{-18}m$. Notice that according to
(\ref{Metric0}) we have that $<\delta\vec{r}>_{(mf)}=0$, but this
result does not imply that $\delta r =0$. A physical consequence
of $\vert<\delta r>\vert\leq 10^{-18}m$ is that any distance
measurement below this value would be meaningless.

Let us add a final comment concerning the relation between
fluctuations and physical effects involved in our paper. Clearly,
different fluctuations can be considered as possible candidates
and several effects can be used as tools in this quest for
precision tests. Our particular choices in this manuscript have
been done looking for the easiest non--trivial cases.
\begin{acknowledgements}
JIRS acknowledges CONACyT grant No. 18176. E. G. thanks UAM--I for
the postdoctoral fellowship.
\end{acknowledgements}

\end{document}